\documentclass{article} 
\usepackage[preprint]{colm2025_conference}

\usepackage{microtype}
\usepackage{hyperref}
\usepackage{url}
\usepackage[most]{tcolorbox}
\usepackage{booktabs}
\usepackage{float}
\usepackage{amsmath,amssymb,amsfonts}
\usepackage{algorithmic}
\usepackage{graphicx}
\usepackage{textcomp}
\usepackage{tabularx}
\usepackage{xcolor}
\newcommand{\correct}{\checkmark}
\newcommand{\xmark}{\texttimes}
\usepackage{todonotes}
\usepackage{comment}
\usepackage{listings}

\definecolor{codegreen}{rgb}{0,0.6,0}
\definecolor{codegray}{rgb}{0.5,0.5,0.5}
\definecolor{codepurple}{rgb}{0.58,0,0.82}
\definecolor{backcolour}{rgb}{0.95,0.95,0.92}

\lstdefinestyle{mystyle}{
  backgroundcolor=\color{backcolour}, commentstyle=\color{codegreen},
  keywordstyle=\color{magenta},
  numberstyle=\tiny\color{codegray},
  stringstyle=\color{codepurple},
  basicstyle=\ttfamily\footnotesize,
  breakatwhitespace=false,         
  breaklines=true,                 
  captionpos=b,                    
  keepspaces=true,                 
  numbers=left,                    
  numbersep=0pt,                  
  showspaces=false,                
  showstringspaces=false,
  showtabs=false,                  
  tabsize=2
}
\usepackage{lineno}

\definecolor{darkblue}{rgb}{0, 0, 0.5}
\hypersetup{colorlinks=true, citecolor=darkblue, linkcolor=darkblue, urlcolor=darkblue}

\title{Semantic Source Code Segmentation using \\ Small and Large Language Models}

\author{Abdelhalim Dahou\textsuperscript{1,*}, Ansgar Scherp\textsuperscript{2}, Sebastian Kurten\textsuperscript{3}, Brigitte Mathiak\textsuperscript{1} \& Madhu Chauhan\textsuperscript{4} \\
\textsuperscript{1} GESIS - Institute for Social Sciences, Cologne, Germany\\
\textsuperscript{2} Ulm University, Ulm, Germany\\
\textsuperscript{3} Utrecht University, Utrecht, Netherlands \\
\textsuperscript{4} IAB - Institut für Arbeitsmarkt- und Berufsforschung, Nürnberg, Germany \\
\texttt{Abdelhalim.dahou@gesis.org} \\
\And
}

\begin{document}

\ifcolmsubmission
\linenumbers
\fi

\maketitle

\begin{abstract}
Source code segmentation, dividing code into functionally coherent segments, is crucial for knowledge retrieval and maintenance in software development. While enabling efficient navigation and comprehension of large codebases, manual and syntactic analysis approaches have become impractical as repositories grow, especially for low-resource languages like R and their research domains (e.g., social sciences, psychology).
This paper introduces an automated, domain-specific approach for research R code segmentation using Large and Small Language Models (LLMs/SLMs). It presents two novel approaches and a human-annotated dataset, StatCodeSeg. We explore two distinct approaches: line-by-line analysis with context and range-based segment determination. 
We experiment with LLMs and fine-tuned SLMs. To support the generalizability of our approaches, we also include experiments on Python code from the computer science domain.
Our results show that context-based line-by-line analysis is superior over range-based segmentation.
Using smaller language models like CodeBERT and an encoder-only version of CodeT5+ are better than their LLM counterparts. 
Most notably, these two best-performing models did not see R code during pre-training versus the LLMs but were only fine-tuned on $4,130$ lines of manually annotated code.
\end{abstract}

\section{Introduction}
\label{sec:introduction}

\begin{figure}[h] 
    \centering 
    \includegraphics[width=1\textwidth]{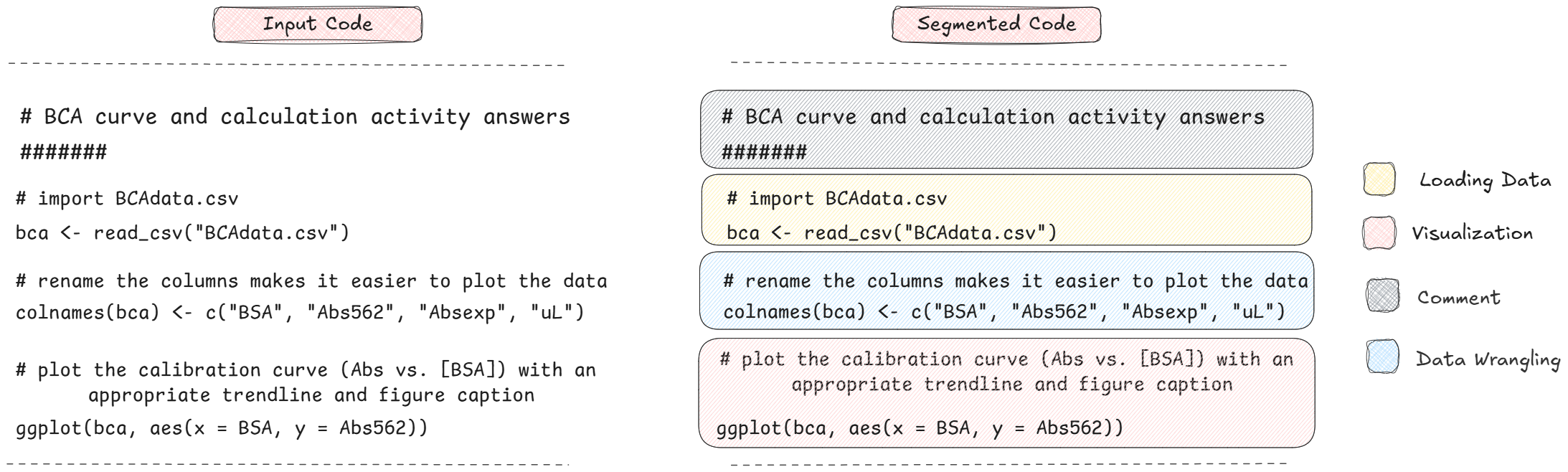}
    \caption{Illustration of code segmentation on example R code,  partitioning code into logical sections  (e.g., Comment, Data Loading, Data Wrangling, and Visualization). \textit{Comment} is a category providing explanatory information.}
    \label{fig:example1}
\end{figure}

The exponential development of code repositories and the rising complexity of programming processes have generated an urgent demand for improved code analysis approaches. Source code segmentation, or the systematic division of code into understandable, logically coherent sections~\cite{stein2022linguistic}, has emerged as a critical task for enhancing code understanding and maintenance, especially in low-resource programming languages like R and their application domains (e.g., social sciences and psychology). In Figure \ref{fig:example1}, we show one example of source code segmentation from an Open Science Framework (OSF) repository\footnote{https://osf.io/rghz3}, where segmentation is based on functionality.

The multifaceted nature of source code, encompassing data processing, statistical analysis, and visualization, makes it challenging to organize and maintain ~\cite{zhang2022coral}. Researchers and users in domains such as social sciences and psychology often rely on less-used programming languages like R, which intensifies these challenges. Although tools like GitHub Copilot can offer assistance, they still depend on robust training data and standardized code practices. Copilot doesn't support R language, widely used by social scientist who ofter share code on platfroms different than Github. Additionally, it struggles with contextual understanding, especially in domain-specific tasks~\cite{bakal2025experience}
We aim to support developers in exploring, analyzing, and managing code by dividing code into logical units. Segmentation may increase code search precision and automated operations like vulnerability detection and documentation generation, increasing code analysis efficiency ~\cite{dormuth2019logical}.

Current methods for determining segment boundaries rely mostly on syntactic structures ~\cite{wang2011automatic}, such as function definitions, control structures, or directly determining a snippet's boundaries through a binary classification task. 
However, logical linkages in code frequently exceed syntactic constraints and notebook cells. 
For example, a data analysis pipeline could have multiple conceptually independent operations such as data preprocessing, wrangling, analysis, and visualization.
All tasks are working toward a common goal but are separated by different syntactic structures and written in different cells ~\cite{zhang2022coral}. 
While many modern Integrated Development Environments (IDEs) effectively manage syntactic links ~\cite{agrawal2024monitor}, identifying logical units within a source code remains crucial for maintaining structure, facilitating updates, and improving documentation. 

Traditional segmentation algorithms fail to incorporate the broader context, which can lead to dividing the data into too many small pieces (fragmented) or segmenting too much together (overly large) that do not reflect logical boundaries. 

The problem remains unresolved due to numerous challenges:  Code's logical boundaries are harder to detect than syntactic structures~\cite{mayer2024evaluating}. Unlike paragraphs in text, code syntax-semantics relationships are not clear. High-quality annotated datasets for logically segmented code are scarce, with existing ones~\cite{dormuth2019logical, stein2022linguistic} only capturing syntactic features. 

LLMs and SLMs present a viable solution to these issues by enabling semantic-aware code segmentation. Furthermore, LLMs like GPT4~\cite{achiam2023gpt} and Code llama~\cite{roziere2023code} can understand the underlying logical structure and functionality of code blocks, particularly within complex analytical workflows, and can support several programming languages~\cite{ma2023lms}. 
This capability may provide a more nuanced understanding of the code’s workflow, resulting in accurate segmentation similar to the segmentation of annotators.
SLMs such as BERT ~\cite{devlin2018bert}, have fewer parameters compared to LLMs but still capture semantic context efficiently through their transformer-based architectures. They offer computationally efficient with faster inference and lower resource consumption.

In this work, we examine code segmentation in R programming as a classification task, employing two approaches: line-by-line and range-based segmentation. The line-by-line approach examines the local context around each code candidate to determine its functionality, whereas the range-based approach uses the complete codebase to construct segment ranges with corresponding functionalities. We implement these approaches using SLMs and LLMs.
We compare the models' performance on StatCodeSeg, a newly created, human-annotated dataset of R code from social science and psychology domains.
Our contributions are in summary:

\begin{itemize}
\item An LM-based approach for code segmentation using SLMs and LLMs with two distinct variants: line-by-line analysis with context-based and range-based.
\item A new human-annotated dataset StatCodeSeg of research R code segments with fine-grained line-based code annotations provided by domain experts.
\item A systematic evaluation of multiple LLMs, including GPT-4o, Gemini 1.5 Pro, Claude 3.5 Sonnet, Code Llama 2, and Qwen2.5-coder, against fine-tuned smaller models CodeBERT and CodeT5+ for code segmentation.
The results reveal that SLMs clearly outperform their large counterparts despite not being exposed to R code during the pretraining phase. 

\item We qualitatively analyze errors and domain-specific challenges in segmenting R code.
\item To support the generalizability of the proposed approaches across domains and programming languages, we extend our evaluation to include additional languages such as Python from computer science domain.

\end{itemize}


\section{Related Work}
\label{sec:relatedwork}

Related works in the area of text segmentation and, particularly, code segmentation are discussed.
Since language models play an important role in our approach, we also discuss the important works on SLM and LLMs in the context of source code.

\subsection{Text Segmentation}
Text segmentation has been the focus of numerous research efforts, employing both supervised and unsupervised neural-based approaches. Early works primarily utilized unsupervised learning methods to quantify lexical cohesion within small text segments. These methods included counting word repetitions ~\cite{choi2000advances} and were later expanded to incorporate lexical chains~\cite{joty2013topic,riedl2012topictiling}. Unsupervised Bayesian approaches were also explored~\cite{mota2019beamseg}, although they encountered challenges in adapting to domain-specific tasks and addressing granularity issues. 
With the advent of BERT~\cite{devlin2018bert}, researchers leveraged modern embeddings to improve performance. Notable examples include BERtSeg ~\cite{solbiati2021unsupervised, park2023unsupervised}, which calculates similarities between utterances to identify topic shifts. On the other hand, supervised algorithms have gained prominence in recent works. 
Many of these studies frame the problem as a sequence labeling task, either by identifying the class of each token~\cite{chauhan2020improving} or determining whether a sentence marks the end of a segment~\cite{koshorek2018text, lukasik2020text, ghosh2023topic}, which often utilize various configurations using BERT, Bi-LSTM and CRF architectures.

In the era of LLMs, advanced text segmentation methods are designed to enhance the logical structure of processed text. LumberChunker~\cite{duarte2024lumberchunker} leverages the reasoning capabilities of LLMs to iteratively predict chunking points, which allows for the effective handling of complex and logically connected texts. However, LumberChunker requires a high level of instruction and computational resources, making it dependent on API-based LLMs, such as Gemini. On the other hand, Meta-Chunking~\cite{zhao2024meta} introduces novel strategies based on LLM to ensure logical coherence while optimizing computational and time efficiency. Despite its strengths, Meta-Chunking requires careful threshold adjustment across datasets, and its adaptability needs further empirical validation.

\subsection{Code Segmentation}
Code segmentation literature is relatively scarce compared to the text segmentation area of research due to the unique challenges posed by languages and the lack of logically segmented code data. 
Similarly to text segmentation, approaches for code segmentation focus on syntax and logical segmentation, leveraging various rules and neural network architectures. Source code segmentation research has evolved from syntax-specific solutions to more language-agnostic approaches. 
The early work by Ning et al.~\cite{ning1994automated} analyzed control flow and data flow in abstract syntax trees (ASTs) of COBOL programs. 
Wang et al.~\cite{wang2011automatic} later proposed generating ASTs for Java methods and applying data flow and syntactic structure rules. Recent studies have explored language-independent methods. 
Dormuth et al.~\cite{dormuth2019logical}  used 
LSTM on a Stack Overflow dataset to segment code based on logical content in six programming languages. 
Stein et al.~\cite{stein2022linguistic} used an LSTM neural network trained on the SCAD dataset from GitHub, using tokenized code snippets and Multi-Segment Code (MSC) blocks to identify segment breakpoints. 
Other studies used code segmentation for Python version identification~\cite{gerhold2024limits} and code summarization~\cite{stein2023chaos}.

\subsection{Code Language Models}

The development of LLMs for code comprehension has experienced significant growth by addressing a wide range of code tasks. This trend began with the introduction of Codex by OpenAI~\cite{chen2021evaluating} and expanded with influential models like CodeGen~\cite{nijkamp2023codegen2}, CodeT5+ ~\cite{wang2023codet5+}, CodeBERT ~\cite{feng2020codebert}, and Starcoder~\cite{lozhkov2024starcoder}. 
Recent models available for download and running locally, such as Code Llama~\cite{roziere2023code}, DeepSeek Coder~\cite{guo2024deepseek}, and Qwen2.5-coder ~\cite{hui2024qwen2} are at the forefront in this field. 
These models, trained on massive code datasets, demonstrated excellent abilities in code understanding and generation tasks. 
Instruction tuning has further broadened their capabilities and improved their performance in various code-related tasks. However, models available via API like GPT-4~\cite{achiam2023gpt}, Claude Sonnet~\cite{anthropic2024claude}, and Gemini~\cite{team2023gemini}~ continue to outperform their lastly mentioned counterparts due to their access to larger and more diverse datasets, greater resources, and optimized architectures.

\section{The StatCodeSeg Dataset}
\label{sec:dataset}

We introduce our StatCodeSeg dataset, its construction process and manually created annotations with validated segment boundaries aligned to code analysis stages. 
Unlike previous approaches like~\cite{dormuth2019logical} that use random concatenation and newline characters as markers (often producing syntactically incorrect combinations), our process ensures both syntactic correctness and meaningful segmentation.

\subsection{Data Preparation}

We selected $160$ files from the StatCodeSearch dataset~\cite{diera2023gencodesearchnet}, comprising $1,070$ code-comment pairs. The original dataset comes from high-quality social science projects collected from OSF \footnote{https://osf.io/} platform. It features R scripts primarily authored by social scientists and psychologists.
The StatcodeSeg was constructed by taking the files directly from the StatCodeSearch dataset and adding additional meta-data features, including license, source, publication/modification dates, etc. 
Before final dataset inclusion, each code file was preprocessed by moving curly/square brackets and parentheses to the end of preceding lines. This simplified input, preserved logical flow, and reduced query count for line-by-line classification.

\subsection{Human Annotation}

We implemented a structured, multi-stage annotation process. We selected $160$ files from the cleaned dataset comprising $13,819$ individual code lines. Three annotators with extensive knowledge of R programming and familiarity with social science and psychology scripts independently labeled each line according to seven predefined code analysis stages. Annotators followed detailed guidelines clarifying the definitions of these stages, as presented in Table \ref{tab:qualitative_rubric} (appendix \ref{appendix:extended_dataset}).

The annotation process was executed in two rounds. Initially, annotators independently code the data, subsequently engaging in a discussion to refine their understanding of the guidelines and specifically address ambiguities related to the \textit{Comment} label and the difference between \textit{Data Wrangling} and \textit{Analysis} stages. 
Following these deliberations, a standardized annotation schema for qualitative coding was developed. This schema clearly defines each stage of data analysis and establishes specific rules for when to apply or avoid each label, and can be found in the Appendix \ref{appendix:extended_dataset}. Using the established annotation schema, the annotators conducted a second independent coding round. 

To measure inter-rater agreement among the three annotators, we used Fleiss' kappa ($\kappa$)~\cite{landis1977measurement}. The obtained score ($\kappa = 0.649$) indicates substantial agreement, reflecting the inherent complexity of the annotation task. The remaining annotation discrepancies were resolved through majority voting, where 95.12\% of lines involved at least two similar categories. In the remaining 4.87\% of cases, a discussion between annotators was conducted to resolve 314 instances arising from \textit{Comment} category ambiguities and $120$ conflicts between other categories, notably, $60$ cases between \textit{Data Wrangling} and \textit{Visualization}, often in scenarios involving data preparation for visualization or styling code.

\subsection{Dataset Characteristics}
The main characteristics of the StatCodeSeg dataset can be found in Table \ref{tab:dataset_stats}. The average number of tokens has been calculated using the CodeBERT tokenizer. Each code line was tokenized using the \textit{tokenize()} method of the tokenizer, and the total number of tokens was averaged across all lines to obtain the final values.

\begin{table}[h!]
\centering
\caption{Statistics of the different subsets used in the StatCodeSeg dataset.}
\label{tab:dataset_stats}

\begin{tabular}{@{}lccc@{}}
\toprule
\textbf{Split} & \textbf{\# Lines} & \textbf{avg Lines/File} & \textbf{avg Tokens/Line} \\
\midrule
Train & 4,130 & 82.60 & 16.95 \\
Val & 782 & 78.20 & 14.16 \\
Test & 8,907 & 89.07 & 18.05 \\
\midrule
StatCodeSeg & 13,819 & 83.29 & 16.83 \\
\bottomrule
\end{tabular}
\end{table}

\section{LM-based Code Segmentation}
\label{sec:methods}

\begin{figure*}[ht]
\centering
\includegraphics[width=\textwidth]{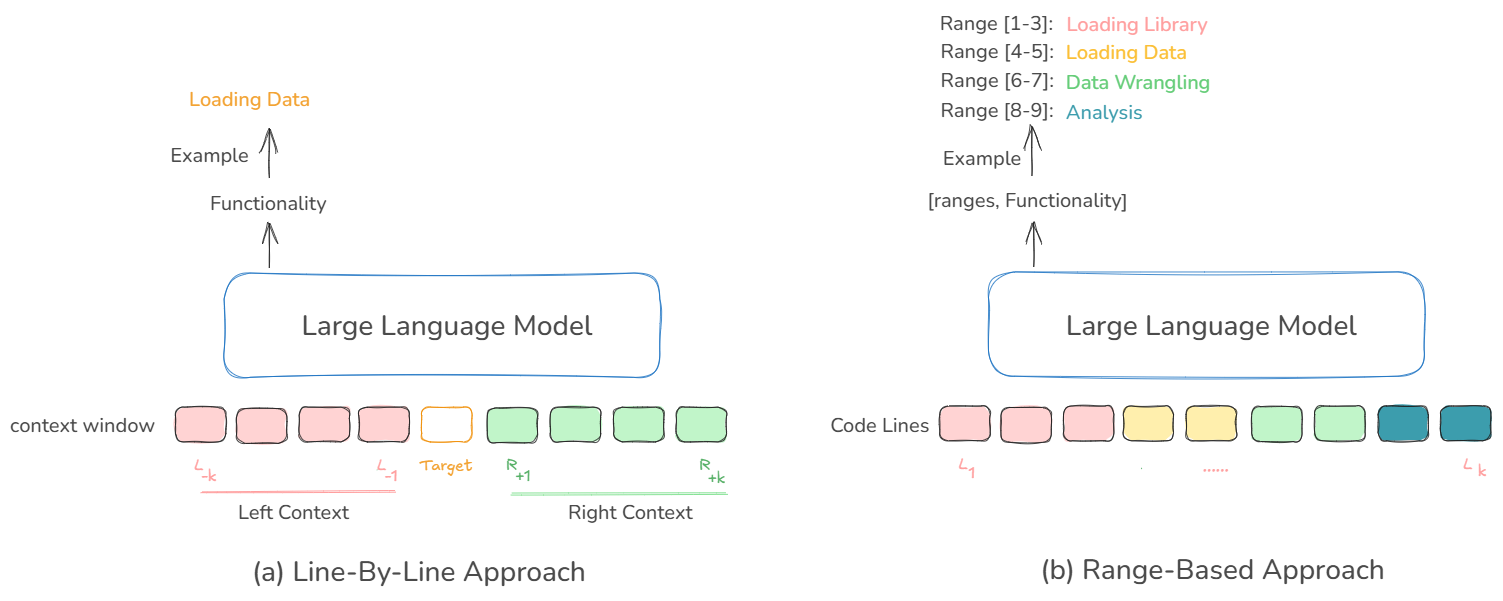}
\caption{Our two proposed segmentation approaches for source code segmentation task. In the line-by-line approach (left), the model is provided with a context window around a target line, consisting of \textit{L} lines to the left and \textit{R} lines to the right. In the range-based approach (right), the model is fed the entire codebase and instructed to generate line ranges, each associated with a specific functionality.}
\label{fig:segmentationapproaches}
\end{figure*}

We describe our code segmentation approach using language models as shown in Figure \ref{fig:segmentationapproaches}.
We have two variants: line-by-line vs. range classification.
\subsection{Line-By-Line Approach}
\label{sec:methods_1}

We propose a two-stage approach for automated code segmentation, focusing first on line-by-line classification (see Figure~\ref{fig:segmentationapproaches}, left) by leveraging the strengths of LLMs, followed by a segment consolidation phase. The initial stage employs a context-window methodology where each line of code is analyzed within its local scope, considering both preceding and following lines. This creates a sliding window effect that moves through the codebase, with each target line being classified according to its functional purpose while taking into account its immediate surroundings. 
The classification process assigns each line to specific code analysis stages, such as \textit{Loading Data, Loading Library}, \textit{Data Wrangling, Analysis}, \textit{Visualization}, \textit{Saving to Output}, or \textit{comment}. 
\begin{verbatim}
<previous_context>
Code_Lines
</previous_context>

<target_line>
Code_Line
</target_line>

<next_context>
Code_Lines
</next_context>
\end{verbatim}

The LLMs are then instructed, via a prompt (see Appendix~\ref{appendix:llmsprompt}), to analyze the input and accompanying code analysis stage descriptions and generate exactly one functional category defining the target line's functionality. A later post-processing phase removed any irrelevant text from the output. 
Using LLMs, our line-by-line technique mimics meta-chunking~\cite{zhao2024meta}, segmenting lines into coherent pieces while considering preceding/following context in various configurations. SLMs, on the other hand, need to be fine-tuned and can tolerate shorter contexts, which leads us to center the target line between equal preceding and following tokens, in cases exceeding model's input tokens, particularly for models like CodeBERT and CodeT5+, which handle only 512 tokens (including two special tokens).  We evaluate the effects of context length on both models using configuration \textit{c} (number of lines before/after target).  Finally, consecutive lines with identical functional categories blend into segments while keeping the sequential order.

\subsection{Range-based Approach}

The range-based approach, as shown in Figure~\ref{fig:segmentationapproaches}, also leverages an LLM to identify functional boundaries within the source code. The model examines the code's structure and functionality to determine appropriate segmentation points, ensuring that each segment represents a complete, standalone code analysis stage. As shown in Figure~\ref{fig:segmentationapproaches}, the LLM processes a code file with multiple lines of code at once and outputs both segments in terms of the range of lines and their corresponding functionality.

The segmentation process begins by receiving a JSON input where each item contains a line number and its corresponding code content, as in  the following example:

\begin{verbatim}
{
{'line': 1, 'code': 'Code_1'}, 
{'line': 2, 'code': 'Code_2'}, 
… 
{'line': k, 'code': 'Code_k'}
}
\end{verbatim}

We employ a zero-shot prompting technique, incorporating functional definitions, reasoning instructions, and desired output format. Although the prompt is comprehensive (see Appendix~\ref{appendix:llmsprompt}).
The model then analyzes the entire codebase to identify functional transitions, outputting ranges such as those shown in Figure~\ref{fig:segmentationapproaches} such as Range [1-3] for Loading Library, Range [4-5] for Loading Data, and so on. This approach ensures that each range encompasses all sequential lines sharing the same functionality, simplifying the understanding of the code's structure and workflow stages.

\section{Experimental Apparatus}
\label{sec:experimentalapparatus}

\subsection{Procedure}
\label{sec:procedure}

To achieve our study objectives, we implement a systematic evaluation approach. 
We begin by preparing two versions of the StatCodeSeg dataset to evaluate different input modalities. Then, we analyze line-by-line versus range-based code segmentation approaches using LLMs by implementing both zero-shot and few-shot prompting techniques. For the few-shot prompting technique, we use the state-of-the-art technique called Clue And
Reasoning Prompting (CARP)~\cite{sun2023text}, more details in Appendix~\ref{appendix:llmsprompt}.
We then compare the performance of LLMs with fine-tuned smaller models through comprehensive evaluation metrics. Throughout all experiments, we maintain consistent evaluation metrics for reliable comparison.

\subsection{Models and Baselines}

\textbf{Baselines.} 
We include two baselines that had not been compared against each other before, as one was proposed for code and the second for text. These are a Bi-LSTM~\cite{stein2022linguistic, mayer2024evaluating} and the Margin Sampling (MSP) of Meta-chunking~\cite{zhao2024meta}, which is the improved method of~\cite{duarte2024lumberchunker}. 
The Bi-LSTM baseline extends the original architecture by transforming it into a multi-class classification task with several context window configurations and customized sentence piece tokenization. The MSP baseline is the state-of-the-art text chunking method uses LLM-based binary classification to determine sentence segmentation based on probabilistic thresholds. The threshold value used in our experiment is 0.5, with the Gemini 1.5 Pro model.

\textbf{Models for Line-by-Line Variant.} 
For the classification task, we use suitable models covering several architectures. 
As an encoder-only model, we use CodeBERT, a bidirectional transformer pre-trained on programming languages that is effective for understanding code context. 
We also include encoder-decoder models like CodeT5+, which combines understanding and generation capabilities. In our experiments, we use the \textit{codet5p-110m-embedding} model variation from HuggingFace. We refer to this version as CodeT5+ (encoder only), as it only comprises the model's encoder layers.
For decoder-only models, we employ Code Llama2 Instruct, which is based on the Llama architecture and fine-tuned for code-specific tasks. Additionally, we leverage advanced LLMs, including Claude 3.5 Sonnet (version 2024-10-22), GPT-4o (version 2024-08-06), Gemini 1.5 Pro, and Qwen2.5-coder, which excel in code-related tasks due to extensive pre-training on both natural language and code data. More details about these models are presented in appendix~\ref{appendix:Modelsdetails}.

\textbf{Models for Range-based Variant.} For this approach, we use models suitable for generation tasks, covering encoder-decoder and decoder-only architectures. In the encoder-decoder category, we use CodeT5+ from HuggigFace called \textit{codet5p-770m}, and for decoder-only architectures, we assess Code Llama 2 Instruct, GPT-4, Claude 3.5 Sonnet, Gemini 1.5 Pro, and Qwen2.5-coder.

\subsection{Measures}
\label{sec:measures}

We evaluate the performance using four multiclass classification metrics: Accuracy, Average Precision/Recall, Micro-F1, and Macro-F1. Macro-F1 provides better assessment for imbalanced classes. If an LLM generates a response that does not match a class, it is counted as a misclassification.

\section{Results}
\label{sec:results}

\label{sec:results-rq1}

Table \ref{tab:performance_metrics_line} shows the evaluation results for the line-by-line classification approach on R programming language (For Python language, see Appendix \ref{appendix:extendedresults}).
The results of each model using different values of the context window (as described in Section~\ref{sec:methods_1})
are illustrated in Figure~\ref{fig:contextwindowimpact} (see Appendix~\ref{appendix:extendedresults}).  
The results show that encoder-only models achieve the highest performance, where CodeBERT achieved the best results with 77.14\% accuracy using a context window of $c=3$. 
CodeT5+ (encoder-only) follows with 74.9\% accuracy with an optimal context of $c=7$. 
Decoder-only models show variable results. Gemini 1.5 Pro achieved $65.65\%$ F1 in zero-shot mechanism and improved performance in few-shot mechanism $70.82\%$ ($c=7$), while Code Llama 2 performed weakest ($45.31\%$ accuracy with $c=1$).
The baselines performed poorly relative to the proposed models, where Bi-LSTM reached $60.89\%$ accuracy ($c=3$).
MSP scored $67.78\%$ accuracy but had the lowest precision $45.22\%$ and recall $44.51\%$.

\begin{table*}
\centering
\caption{Results for Line-by-Line Classification on R Language.}
\label{tab:performance_metrics_line}
\begin{tabular}{lcccccc}
\toprule
Model & Acc. & Precis. & Recall & Macro F1 & Micro F1 & Context\\ 
\midrule
\textit{Baselines} \\
Bi-LSTM & 60.89 & 68.6 & 58.85 & 61.48 & 61.73 & 3\\
Meta-Chunking (MSP) & 67.78 & 45.22 & 44.51 & 43.09 & 67.78 & -\\
\midrule
\textit{Fine-tuned Models} \\
CodeBERT & \textbf{77.14} & \textbf{79.72} & 79.87 & \textbf{79.49} & \textbf{77.14} & 3\\
CodeT5+ (encoder only) & 74.9 & 75.76 & 76.95 & 76.2 & 74.9 & 7\\
\midrule
\textit{Zero-shot Models} \\
Claude 3.5 Sonnet & 69.92 & 62.26 & 63.92 & 61.1 & 69.92 & 3 \\
GPT-4o & 64.52 & 61.51 & 59.48 & 57.42 & 64.52 & 2 \\
Gemini 1.5 Pro & 75.92 & 64.23 & 69.32 & 65.65 & 75.92 & 7 \\
DeepSeek V3 & 74.51 & 75.43 & \textit{77.72} & 75.03 & 74.51 & 1 \\
DeepSeek R1 & \textit{76.53} & \textit{79.03} & \textbf{81.50} & \textit{78.87} & \textit{76.53} & 2 \\
Qwen2.5-coder & 69.37 & 26.51 & 29.07 & 27.02 & 69.37 & 3 \\
Code Llama 2
& 45.31 & 07.56 & 07.18 & 06.89 & 45.31 & 1\\ 
\midrule
\textit{Few-shot Models (CARP \textsubscript{16-shot})} \\
Claude 3.5 Sonnet & 61.60 & 15.28 & 19.59 & 16.12 & 61.60 & 3 \\
GPT-4o & 63.47 & 66.48 & 70.20 & 65.09 & 63.47 & 2 \\
Gemini 1.5 Pro  & 68.82 & 72.35 & 74.89 & 70.82 & 68.82 & 7 \\
DeepSeek V3 & 51.40 & 36.49 & 20.10 & 22.62 & 51.40 & 1 \\
Qwen2.5-coder & 30.47 & 16.39 & 13.07 & 12.62 & 30.47 & 3 \\
Code Llama 2 & - & - & - & - & - & - \\
\bottomrule
\end{tabular}
\end{table*}

\label{sec:results-rq2}
Results of the range-based evaluation approach are provided in Table~\ref{tab:performance_metrics_range_}. Claude 3.5 Sonnet achieved the highest accuracy of $76.81\%$ among decoder-only models. Gemini 1.5 Pro followed with $75.2\%$ accuracy, while GPT-4o achieved $70.91\%$ accuracy. Encoder-decoder CodeT5+ underperformed with $24.61\%$ accuracy, and Code Llama 2 produced invalid outputs, which makes it hard to evaluate for this approach. Compared to the line-by-line approach (Table~\ref{tab:performance_metrics_line}), decoder-only models such as Claude 3.5 Sonnet ($76.81\%$ vs. $69.92$\%) and Gemini 1.5 Pro ($75.2\%$ vs. $76.99\%$) show comparable accuracy in the range-based method. 

\begin{table*}[t]
\centering
\caption{Results for Range-Based on R Language.}
\label{tab:performance_metrics_range_}
\begin{tabular}{lccccc}
\toprule
Model & Acc. & Precision & Recall & Macro F1 & Micro F1\\ 
\midrule
\textit{Encoder-decoder Models} \\
CodeT5+ & 24.61 & 04.44 & 02.91 & 03.36 & 24.61 \\
\midrule
\textit{Decoder-only Models} \\
Claude 3.5 Sonnet & \textbf{76.81} & \textbf{62.76} & \textbf{65.89} & \textbf{63.96} & \textbf{76.81}  \\
GPT-4o & 70.91 & 44.79 & 45.93 & 44.96 & 70.91  \\
Gemini 1.5 Pro & \textit{75.2} & \textit{62.55} & \textit{64.43} & \textit{62.7} & \textit{75.2}  \\
DeepSeek V3 & 73.20 & 62.17 & 63.41 & 61.57 & 73.20  \\
DeepSeek R1 & 71.69 & 59.56 & 56.48 & 56.11 & 71.69  \\
Qwen2.5-coder & 60.85 & 13.48 & 13.34 & 13.19 & 60.85  \\
Code Llama 2 & - & - & - & - & - \\
\bottomrule
\end{tabular}
\end{table*}


\section{Discussion}
\label{sec:discussion}




%
Encoder-only models such as CodeBERT demonstrate superior performance in classification through effective local context capture, followed by decoder-only models, in particular, Gemini 1.5 Pro model using few-shot mechanism (Figure~\ref{fig:promptimpact}, Appendix~\ref{appendix:extendedresults}). LLMs such as Claude 3.5 Sonnet and Gemini 1.5 Pro excel in range-based segmentation due to their expanded context windows and understanding/generative capabilities. This architectural distinction extends to performance trade-offs, where smaller models remain competitive in line-by-line tasks but face limitations in complex, context-heavy segmentation. The impact of context window optimization is evident in the results, with CodeBERT achieving optimal classification performance at $c=3$, while CodeT5+ (encoder-only) required $c=7$, demonstrating how architectural differences and training strategies influence optimal context size configurations.

CodeBERT achieves the highest accuracy ($77.14\%$) in the classification approach with a moderate local context, surpassing even larger decoder-only models such as GPT-4o ($64.52\%$ accuracy). This aligns with previous findings~\cite{feng2020codebert,wang2023codet5+,diera2023gencodesearchnet}
%
%
that code-focused encoder-only models have been shown to achieve high performance in classification tasks, where fine-grained syntactic and semantic patterns in code are crucial. Comparing zero-shot and fine-tuning, we can observe that zero-shot prompting with LLMs can yield highly competitive (even superior in range-based approach) results in segmenting code based on functionality. This reduces the overhead of fine-tuning while still capturing functionality transitions, except for complex codebase, as shown by~\citet{ma2023lms}, that LLMs (GPT-4, Code Llama) are strongly affected by the data-shift problem and lack in-depth reasoning capabilities about the data flow. 

However, fine-tuned models remain effective in this task (with training data), as seen with CodeBERT and CodeT5+ (encoder-only). For example, in our analysis (see Figure~\ref{tab:segment_counts}, Appendix~\ref{appendix:extendedresults}), CodeBERT demonstrated a much lower average difference in terms of the number of segments ($7.11$) from annotator defined segmentation compared to the best LLM (Claude 3.5 Sonnet) in range-based approach ($20.55$), reflecting its ability to closely align with expert segmentation in files that have higher complexity in terms of functionality shifting.


\section{Conclusion and Future Work}
\label{sec:conclusion}
This paper introduces two approaches based on SLMs and LLMs for the source code segmentation task. A line-by-line approach uses only local context (preceding and following) around a target line to be classified, and a range-based approach treats a full codebase and generates segments with code analysis functionality. We evaluated these two approaches on a new human-annotated dataset called StatCodeSeg of R code segments with fine-grained line-based code annotations done by domain experts.
We compared our approaches with recent baselines based on neural approaches.

Our experiments in both approaches surpassed baseline models. Encoder-only models (CodeBERT) excel in classification through local context capture, while LLMs like Claude and Gemini perform better in range-based segmentation due to expanded context windows. Fine-tuned encoder-only models outperform LLMs even without seeing R code during pre-training, using only $4,130$ lines of annotated code. Few-shot prompting improves LLM classification results compared to zero-shot, though limitations persist in complex codebases due to data shift and reasoning constraints.

In future work, we plan to explore the impact of various prompting techniques in the context of LLMs. Additionally, incorporate data augmentation strategies to expand the dataset, thereby enhancing model generalization.

\newpage

\bibliography{colm2025_conference}
\bibliographystyle{colm2025_conference}

\appendix

\section{Supplementary Materials}
\label{appendix:supplementarymaterials}
\subsection{Model Details}
\label{appendix:Modelsdetails}

Table \ref{tab:code_models} compares the used code models in terms of input modalities, training objectives, model sizes, and programming languages seen during training.
\begin{table*}[!ht]
    \centering
        \caption{Comparison of code models in terms of input modalities, training objectives, model sizes, and programming languages seen during training.  Key objectives include RTD (identifying replaced tokens), MLM (predicting masked tokens), IT (classifying tokens as identifiers), BDG (generating NL summaries or code), FIM (filling in missing code), and NTP (predicting the next token).}
    \label{tab:code_models}
    \begin{tabular}{@{}lcccccc@{}}
        \toprule
        \textbf{Models} & \textbf{Input} & \textbf{Objectives} & \textbf{Param.} & \textbf{Progr.} & \textbf{R} \\
        \midrule
        \textit{Encoder Models} \\
        CodeBERT                & Code + Doc & MLM \& RTD    & 125M    & 6  & \xmark \\
        CodeT5+                  & Code + Doc & MLM / IT / BDG & 110M    & 7  & \xmark \\
        \midrule
        \textit{Decoder-only Models} \\
        Code Llama 2 Instruct      & Code + Doc & Infilling      & 7B      & 7  & \xmark \\
        GPT-4o                & Code + Doc & NTP            & --      & -- & \correct \\
        Claude 3.5 Sonnet       & Code + Doc & NTP            & --      & -- & \correct \\
        Gemini 1.5 Pro          & Code + Doc & NTP            & --      & -- & \correct \\
        Qwen2.5-coder              & Code + Doc + Math      & FIM, NTP           & 7B   & 92 & \correct \\

        DeepSeek V3            & Code + Doc + Math      & MTP, RLHF           & 37B   & - & - \\
        \bottomrule
    \end{tabular}

\end{table*}

\subsection{Hyperparameter Optimization}
\label{appendix:hyperparameteroptimization}
We tuned CodeBERT's hyper-parameters on the validation set, exploring learning rates of $[1-5] \times 10^{-5}$, batch sizes [$8$, $16$, $32$, $64$], weight decay [$0.01$ - $0.03$], and epochs [$2$ - $8$], before applying the optimal values, learning rate of $3 \times 10^{-5}$, batch size of $16$,  epochs $8$,  weight decay of $0.02$. 
For CodeT5+ models, we maintain their original parameters while setting the sequence length to $512$ tokens, with batch sizes of $8$ and $1$ for encoder-only and encoder-decoder variants, respectively, and $100$ epochs for the latter. 
For LLMs, we use the standard hyper-parameters.

\subsection{LLMs Prompts}
\label{appendix:llmsprompt}
Below is the prompt template used to evaluate the LLMs in both approaches. For the few-shot prompting technique, we use a state-of-the-art technique called Clue And
Reasoning Prompting (CARP). This method first identifies relevant clues in the input text, such as keywords, functions, libraries, contextual information, and data structures. Then develops a comprehensive reasoning process that goes beyond superficial pattern matching to determine the correct functionality. Examples are strategically sampled using CodeBERT fine-tuned embeddings to provide relevant demonstrations with appropriate clues and reasoning information.

\begin{tcolorbox}[colback=gray!10!white, colframe=black, title=Line-by-Line classification Prompt, fonttitle=\bfseries, sharp corners=southwest, enhanced, breakable, boxrule=0.5pt, width=\textwidth, arc=2mm, listing only, listing options={style=tcblatex, basicstyle=\ttfamily\small}]

 \# Objective:\\
Classify R programming code line of a  given source code into one of the next 
categories based on its function:\\
\{All labels with description\}

\#Rules: \\
\{Rules to follow\}

\#Note:\\
    OUTPUT: JUST THE CATEGORY LABEL 
    WITHOUT ANY ADDITIONAL TEXT.
    \\
\{Example\}
\end{tcolorbox}

\begin{tcolorbox}[colback=gray!10!white, colframe=black, title=Range-based Prompt, fonttitle=\bfseries, sharp corners=southwest, enhanced, breakable, boxrule=0.5pt, width=\textwidth, arc=2mm, listing only, listing options={style=tcblatex, basicstyle=\ttfamily\small}]
You are an AI assistant specialized in categorizing R  source code by functionality 
in contiguous line ranges.

You receive the entire source code as a JSON array of objects, where each object contains:

    - "Line": the line number (integer)
    
    - "Code": the text of that line 
        (string)

Your task is to:

1. Classify the lines into the following categories:
\{All labels with description\}

2. Merge consecutive lines of the *same* category into a single range.

3. Output only the range-based  classification for the entire file,  with no additional 
text or explanation.

\{Important Details\}

Your final answer must be strictly the ranges in plain text, nothing else. 

\end{tcolorbox}

\subsection{Extended Results}
\label{appendix:extendedresults}

\begin{table*}[h!]
\centering
\caption{Comparative performance of models using Line-by-Line classification approach for Python language.}
\label{tab:performance_metrics_line}
\begin{tabular}{lcccccc}
\toprule
Model & Acc. & Precis. & Recall & Macro F1 & Micro F1 & Context\\ 
\midrule
\textit{Fine-tuned Models} \\
CodeBERT & \textit{85.59} & \textbf{90.67} & \textit{81.76} & \textbf{84.39} & \textit{85.59} & 3\\

CodeT5+ (encoder only) & \textbf{85.62} & \textit{89.96} & 78.38 & \textit{79.71} & \textbf{85.62} & 7\\
\midrule
\textit{Zero-shot Models} \\
Claude 3.5 Sonnet & 79.94 & 54.47 & 76.44 & 58.51 & 79.94 & 3 \\
GPT-4o & 71.51 & 63.84 & \textit{83.39} & 64.58 & 71.51 & 2 \\
Gemini 1.5 Pro & 81.01 & 63.64 & \textbf{89.47} & 68.68 & 81.01 & 7 \\
DeepSeek V3 & 81.13 & 65.27 & 85.02 & 67.69 & 81.13 & 3 \\
DeepSeek R1 & 79.43 & 62.67 & 83.82 & 66.12 & 79.43 & 2 \\
Qwen2.5-coder & 67.08 & 41.52 & 58.05 & 42.35 & 67.08 & 3 \\
Code Llama 2
& 34.10 & 23.87 & 28.77 & 18.61 & 34.10 & 1\\ 
\bottomrule
\end{tabular}
\end{table*}

\begin{table}[H]
    \centering
    \caption{Comparison of segmentation counts and model performance using MAE (Mean Absolute Error) and STD (Standard Deviation). Comparison is across selected R source code files that exhibit significant differences in the number of segments between annotators and models.}
    \begin{tabular}{cccc}
        \toprule
        \textbf{ID} & \textbf{Annotators} & \textbf{Claude 3.5 Sonnet} & \textbf{CodeBERT} \\
        \midrule
        5  & 43 & 15  & 16   \\
        29 & 37 & 14  & 26  \\
        52 & 29 & 12  & 27  \\
        74 & 3  & 34  & 5   \\
        77 & 16 & 33  & 20  \\
        81 & 28 & 12  & 36  \\
        85 & 27 & 11  & 25  \\
        93 & 38 & 20  & 34  \\
        97 & 2  & 21  & 6    \\
        \midrule
        \textbf{MAE} & --  & 20.56  & \textbf{7.11} \\
        \textbf{STD} & --  & 5.55 & 8.05  \\
        \bottomrule
    \end{tabular}
    \label{tab:segment_counts}
\end{table}

\begin{figure}[ht]
\centering
\includegraphics[width=0.8\textwidth]{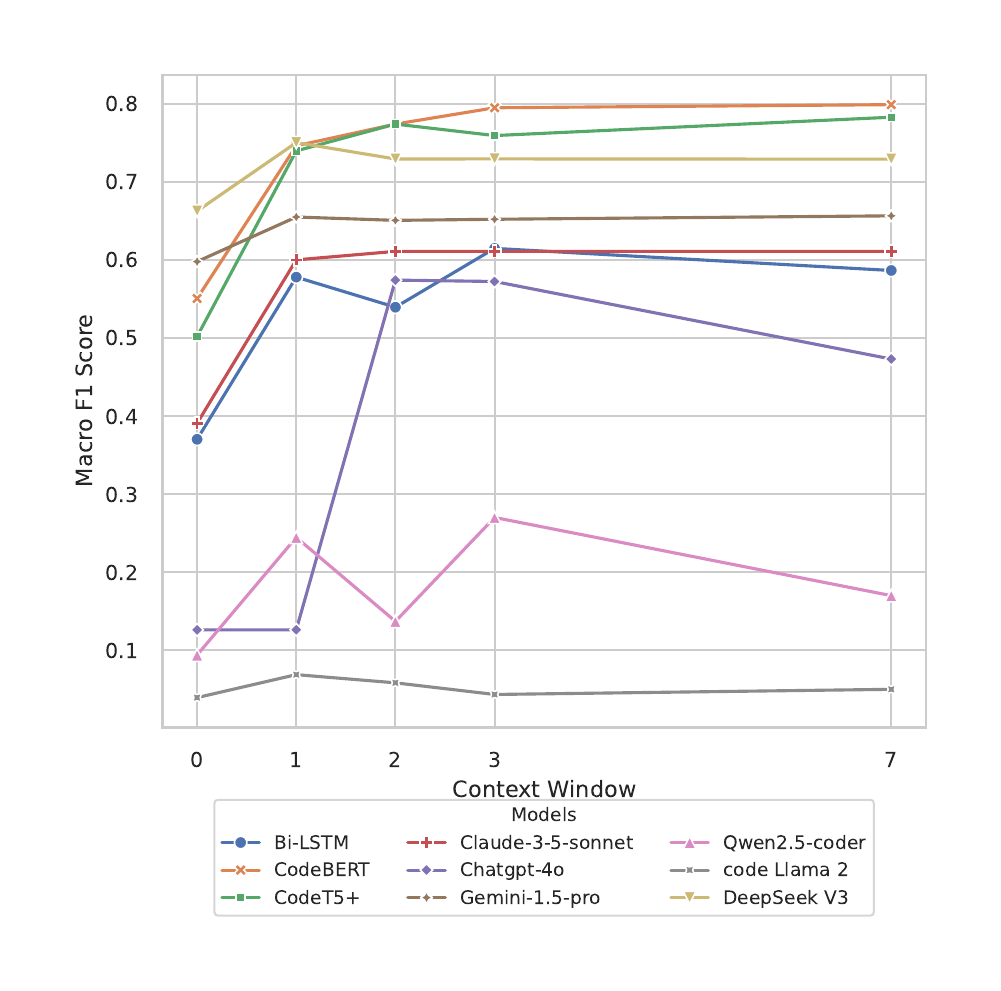}
\caption{Analysis of the importance of the size of context window in the Line-by-Line approach for R language.}
\label{fig:contextwindowimpact}
\end{figure}

\begin{figure}[ht]
\centering
\includegraphics[width=0.80\textwidth]{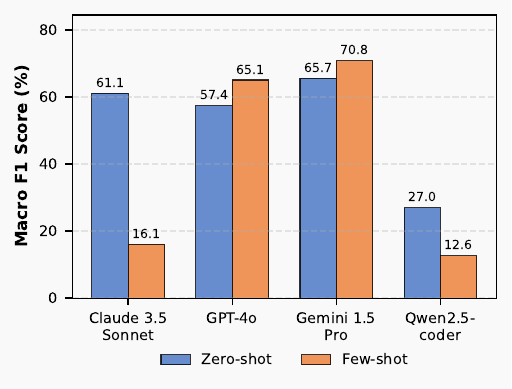}
\caption{Comparative analysis of zero-shot versus few-shot prompting techniques for LLMs in the Line-by-Line classification approach for R language.}
\label{fig:promptimpact}
\end{figure}

\subsection{Limitations}
\label{appendix:limitations}
This work introduces a novel dataset, though, like many dataset creation efforts, there is a potential data leakage, where parts of the StatCodeSeg test set might have been included in the model’s pre-training data. However, as noted in ~\cite{diera2023gencodesearchnet}, since the dataset derives from OSF rather than public platforms such as GitHub, we assume that existing language models were not exposed to this data during pre-training. In addition, StatCodeSeg is currently limited to R programming language and contains fewer training samples compared to the test set. Despite the widespread use of alternative tools like SPSS, STATA, SAS, and Python in the social science and psychology domains. To enhance the coverage of the dataset  across other domains and support different programming languages, future iterations of StatCodeSeg could incorporate these features.

As shown in Table \ref{tab:performance_metrics_range_}, we did not report the results of Code Llama 2 using the range-based approach. Although the model was applied to this study (as shown in Table \ref{tab:performance_metrics_line}), it consistently produced incorrect outputs that lacked any ranges or functionality annotations. In most runs, the model generated a full textual description of the input code instead than following the provided instructions, making it unsuitable for evaluation.

\subsection{Threat to Validity}
\label{sec:threattovalidity}
The first concern is related to the way prompts are designed in this study. To minimize disruption to the LLM's performance, we meticulously craft and refine the prompt template, incorporating clear explanations of tasks and concepts. To ensure the desired output, the prompts were initially generated using GPT-4o’s developer mode (specialized for prompt creation) and iteratively refined through a combination of automated refinement (via Claude’s interface) and manual adjustments based on model responses and a process repeated over three runs. Additionally, the template explicitly separates output examples from the input section to help models distinguish classification targets from example cases. 

The second issue involves the size of the context window of the classification approach. 
A small input length could fail to capture longer dependencies or relationships between code lines separated by significant distances.
However, as we already see from our analysis of the influence of context sizes (see Figure \ref{fig:contextwindowimpact}), either the model performance levels off with higher context size or even harms the performance.
Thus, it seems reasonable that the experiments identified optimal context window sizes.
With an average number of $16.83$ tokens per code line (see Table \ref{tab:dataset_stats}), the input fits well even in the limited input size of $512$ like CodeBERT.

\subsection{Annotation Conflicts}
\label{appendix:detaileddiscussion}

After the second round of test set annotation process, we observed conflicts in 434 cases, representing 4.87\% of the total annotations. Most of these conflicts (72.35\%, 314 cases) were related to comment categorization issues, which were resolved manually by taking care of the adjacent code line. The remaining 120 cases presented more substantial categorical disagreements among the annotators.\\
The most frequent conflicts occurred between the data wrangling and visualization categories (60 cases), followed by conflicts between data wrangling and analysis (20 cases). We also identified 13 instances where annotators disagreed between loading data and data wrangling classifications. Less frequent conflicts included disagreements between analysis and comments (5 cases), analysis and visualization (4 cases), and data wrangling and comments (3 cases). Additionally, we observed 9 cases of conflict between saving outputs and data wrangling, and 3 cases each for conflicts between saving outputs and loading data.

\subsection{Annotaiton Rules}
\label{appendix:extended_dataset}
Table \ref{tab:qualitative_rubric} presents detailed definitions for each category and the labeling rules employed by the annotators during the second round of annotation for the StatCodeSeg dataset. An example is provided to clarify and illustrate each category.

\begin{table*}[!ht]
\centering
\caption{Definitions, labeling rules, and examples for each code analysis category provided to annotators for labeling the StatCodeSeg dataset.}
\begin{tabular}{p{1.9cm}|p{4cm}|p{4.1cm}|p{4.5cm}}
\hline
\textbf{Stage} & \textbf{Definition} & \textbf{When to Use} & \textbf{Example} \\
\hline
Loading \\ Library & 
Loads an external library that provides specific functionalities. & 
Loading packages, setting directory, using 'Library' keyword & 
\lstset{style=mystyle}
\begin{lstlisting}[language=R]
 library(mgcv)
 library(itsadug)
 library(sjPlot)
\end{lstlisting} \\
\hline
Loading \\ Data & 
Loads data from an outside source into the R environment. & 
Reading data from files or databases. If "Loading Data" appears most frequently across all lines of code, label the entire snippet as "Loading Data". &
\lstset{style=mystyle}
\begin{lstlisting}[language=R]
 read_data<- function(data_path) {
 dat <- read_csv(data_path,
 col_types = cols())
\end{lstlisting} \\
\hline
Data \\ Wrangling & 
Shifts data around without adding new information. This includes data transformation such as cleaning, filtering, summarizing, and augmenting an existing dataset. & 
When the code involve direct
manipulation of data structures to make data more accessible or manageable.& 
\lstset{style=mystyle}
\begin{lstlisting}[language=R]
 # join id-number to predicted data
 predt<-cbind(test$id,pred)
 colnames(predt)<-c("id", 
    "class_var_pred")
\end{lstlisting} \\
\hline
Analysis & 
Train and evaluate models to learn relationships in the data, perform statistical analysis, and generate insights for predictions or data understanding. & 
Statistical modeling, fitting and/or specifying machine learning models, simulation, and defining loss functions. & 
\lstset{style=mystyle}
\begin{lstlisting}[language=R]
 return_Cohen_d <- function(var_1){
 mean_var1 <- mean(var_1)
 sd_var1 <- sd(var_1)
 Cohen_d <- mean_var1 / sd_var1
 return(Cohen_d) }
\end{lstlisting} \\
\hline
Visualization & 
Creates visual outputs of data, such as writing to an image file or displaying on the screen. & 
Any code that defines or adjusts visual properties for data display, like color
schemes or formatting, should be classified as "Visualization." & 
\lstset{style=mystyle}
\begin{lstlisting}[language=R]
 virus_colors <- list(
 HCoV-19 = "#cc4747",
 SARS-CoV-1 = "#7570b3", 
 MERS-CoV = "#518591" )
\end{lstlisting} \\
\hline
Saving \\To Output & 
Saves data or results to a file, such as a CSV or text file. & 
Writing analysis results to a file using functions like write.csv(), write.table(), or saveRDS() to save data frames or objects to external files. & 
\lstset{style=mystyle}
\begin{lstlisting}[language=R]
 png("output_plot.png")
 plot(my_data)
 dev.off()
\end{lstlisting} \\
\hline
Comment & 
Provides additional information that is different from the next code line, explaining the process or the experiment/task. & 
When a comment line (\textcolor{green}{\#}) does not describe the functionality of the subsequent lines of code, categorize it simply as 'Comment'.& 

\lstset{style=mystyle}
\begin{lstlisting}[language=R]
 # Success predictors -----------
 model.fit(X, y)
\end{lstlisting} \\
\hline
\end{tabular}
\label{tab:qualitative_rubric}
\end{table*}

\end{document}